\documentclass[aps,prl,twocolumn,groupedaddress,showpacs,showkeys]{revtex4-1}
\usepackage{graphicx}
\usepackage{array}
\usepackage{SIunits}
\usepackage{color} % %
\begin{document}

% Use the \preprint command to place your local institutional report
% number in the upper righthand corner of the title page in preprint mode.
% Multiple \preprint commands are allowed.
% Use the 'preprintnumbers' class option to override journal defaults
% to display numbers if necessary
%\preprint{}

%Title of paper
\title{Spectral engineering of slow light, cavity line-narrowing and pulse compression}

% repeat the \author .. \affiliation  etc. as needed
% \email, \thanks, \homepage, \altaffiliation all apply to the current
% author. Explanatory text should go in the []'s, actual e-mail
% address or url should go in the {}'s for \email and \homepage.
% Please use the appropriate macro foreach each type of information

% \affiliation command applies to all authors since the last
% \affiliation command. The \affiliation command should follow the
% other information
% \affiliation can be followed by \email, \homepage, \thanks as well.
\author{Mahmood Sabooni, Qian Li, Lars Rippe, R. Krishna Mohan*, Stefan Kr\"{o}ll}
%\email[]{Your e-mail address}
%\homepage[]{Your web page}
%\thanks{}
%\altaffiliation{}
\affiliation{Department of Physics, Lund University, P.O.~Box 118, SE-22100 Lund, Sweden}
\affiliation{*Spectrum Lab, P. O. Box 173510, Montana State University, Bozeman, Montana 59717}
%Collaboration name if desired (requires use of superscriptaddress
%option in \documentclass). \noaffiliation is required (may also be
%used with the \author command).
%\collaboration can be followed by \email, \homepage, \thanks as well.
%\collaboration{}
%\noaffiliation

%\date{\today}

\begin{abstract}
More than four orders of magnitude cavity-linewidth narrowing in a rare-earth-ion-doped crystal cavity, emanating from strong intra-cavity dispersion caused by off-resonant interaction with dopant ions is demonstrated. The dispersion profiles are engineered using optical pumping techniques creating significant semi-permanent but reprogrammable changes of the rare earth absorption profiles. Several cavity modes are shown within the spectral transmission window. Several possible applications of this phenomenon are discussed.
\end{abstract}

% insert suggested PACS numbers in braces on next line
% insert suggested keywords - APS authors don't need to do this
%\keywords{}

%\maketitle must follow title, authors, abstract, \pacs, and \keywords
\maketitle

% body of paper here - Use proper section commands
% References should be done using the \cite, \ref, and \label commands
%\section{Introduction}
% Put \label in argument of \section for cross-referencing
%\section{\label{}}
Cavity linewidth narrowing has been suggested to have great potential in many different areas such as laser stabilization \cite{Shevy2010a,Lukin1998}, high-resolution spectroscopy \cite{Lukin1998}, enhanced light matter interaction and compressed optical energy \cite{Baba2008}. We show more than four orders of magnitude cavity linewidth narrowing, which, to the best of our knowledge, is more than two orders of magnitude larger than demonstrated with other techniques. Previously 10 to 20 times linewidth narrowing has been shown using either EIT \cite{Wu2008,Zhang2008,Wang2000a}, and recently two orders of magnitude was shown using coherent population oscillation (CPO) in combination with a cavity dispersive effect \cite{Grinberg2012a}. We also demonstrate several cavity modes within the slow light transmission window, something which we are not aware of having been demonstrated using EIT or CPO. The present results are obtained using spectral hole-burning in rare-earth-ion doped crystals \cite{Shakhmuratov2005,Lauro2009,Walther2009a} and we discuss the properties and potential of slow light structures created with this method in these materials.

\par
In this paper, a cavity formed by depositing mirrors directly onto a praseodymium doped $Y_2SiO_5$ crystal and (near) persistent spectral hole burning (PSHB) is employed to create a very strong dispersion. A sharp dispersion slope reduces the photon group velocity, and therefore increases the effective photon lifetime in the cavity compared to a non-dispersive cavity (some times referred to as cold cavity \cite{Siegman1985}).
\par
Generally the mode spacing in a Fabry-P\'{e}rot cavity, $\Delta\nu$, is given by \cite{Siegman1985}
\begin{equation}\label{eq:Mode_Spacing}
    \Delta\nu=\frac{c}{2L}\frac{1}{n_g(\nu)}=\frac{c}{2L}\frac{1}{n+\nu\frac{dn}{d\nu}}=\frac{v_g(\nu)}{2L}
\end{equation}
where c is the speed of light in vacuum, $\nu$ is the light frequency, $n$ is the real part of the index of refraction (for the phase velocity), $v_g(\nu)$ is the group velocity and $n_g(\nu)$ is the index of refraction for the group velocity. For the present work it is useful to briefly analyse the mode spacing relation.

\par
The resonance condition for a Fabry-P\'{e}rot cavity of length L may be expressed as $m(\lambda/2)=L$, where $m$ is an integer and the wavelength $\lambda=c/(n \nu)$. Thus
\begin{equation} {\label{eq:cmodes}}
m \frac{c}{2L} = n \nu
\end{equation}
Differentiating Eq. \ref{eq:cmodes} gives
\begin{equation} {\label{eq:dcmodes}}
\frac{c}{2L} \delta m = n \delta \nu + \nu \delta n
\end{equation}
Dividing the left (right) hand side of Eq. \ref{eq:dcmodes} with the left (right) hand side of Eq. \ref{eq:cmodes} yields
\begin{equation} {\label{eq:dcmodes2}}
\frac{\delta m}{m} = \frac{\delta \nu}{\nu} + \frac{\delta n}{n}
\end{equation}

\begin{figure}[ht]
    \includegraphics[width=8cm]{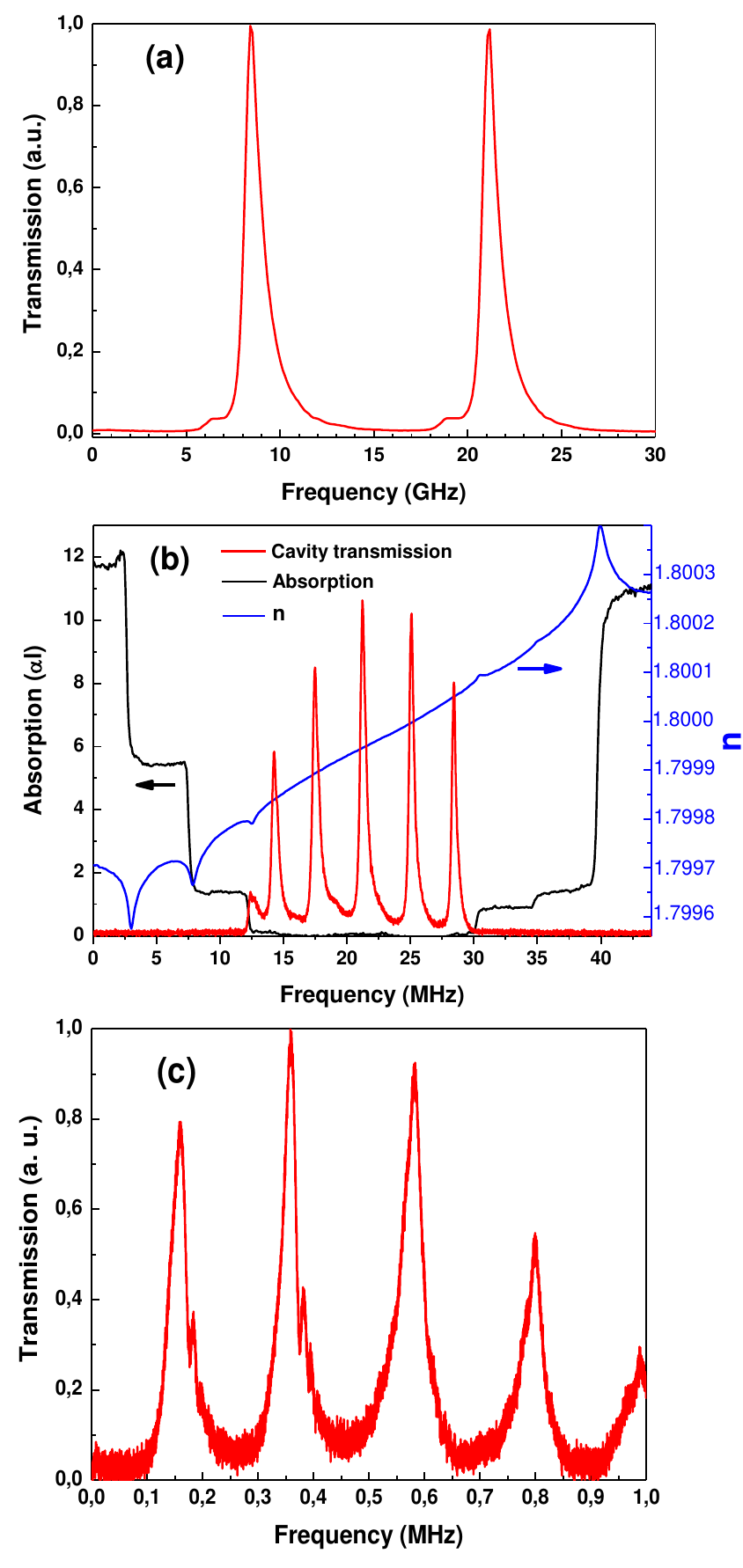}
    \caption{(Color online) (a) Non-dispersive cavity transmission spectrum. (b) Cavity transmission spectrum (red trace), best guess for the absorption profile around the $\approx18\;$MHz spectral hole (black trace) inside the cavity. The corresponding real refractive index, $n$, calculated from the black trace absorption profile is shown as a blue trace. (c) Cavity transmission spectrum for the absorption profile about a MHz spectral hole.}
    \label{Readout}
\end{figure}
where $n = n(\nu)$ is a function of frequency. Normally when the frequency is changed $\delta \nu/\nu \gg \delta n/n$, but in the case of significant slow light effects $ n \ll \nu (d n / d\nu)$ and the second term on the right hand side in Eq. \ref{eq:dcmodes2} is much larger than the first. Thus the cavity mode spacing is basically completely determined by the dispersion while the impact of the relative change in the frequency is negligible. Below it is shown how we modify the cavity absorption to enter this regime where the Fabry-P\'{e}rot cavity mode spacing is completely determined by the frequency dispersion.

\par
\begin{figure}[ht]
    \includegraphics[width=8cm]{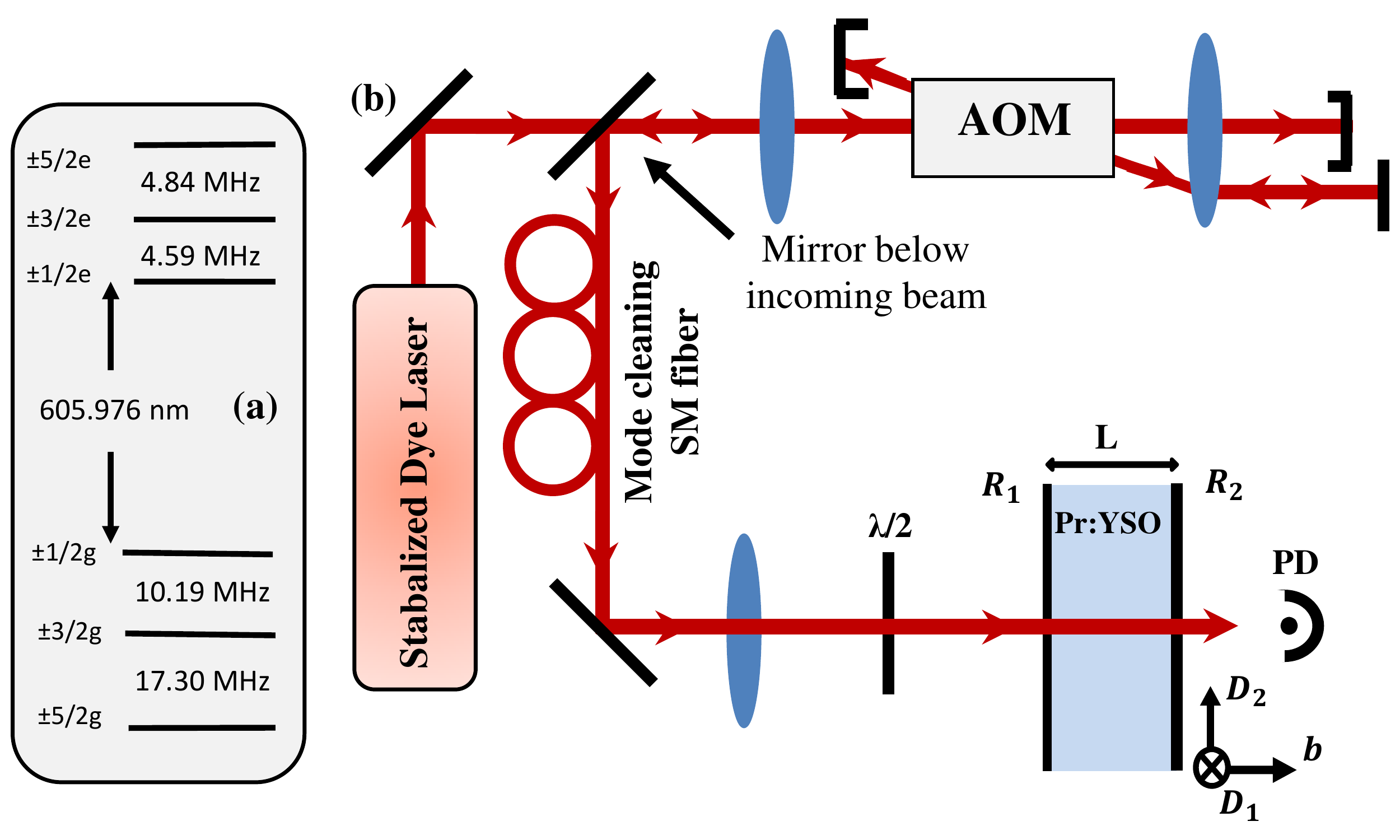}
    \caption{(Color online) Experimental set up. (a) The hyperfine splitting of the $^3H_4-\!^1D_2$ transition of site I $Pr^{3+}\!:\!Y_2SiO_5$ is shown \cite{Equall1995,Rippe2005}. (b) A double pass acousto-optic modulator (AOM) is employed to tailor the optical pulses out of the continuous output from a narrow ($<\,$kHz) linewidth laser. A photo diode, PD, monitors the light transmitted through the cavity. The specified reflectivity of the crystal input and output facets is $R_1=R_2=95\%$. $D_1$, $D_2$ and $b$, show the crystal principal axes orientations.}
    \label{setup}
\end{figure}

First, we examine the cavity transmission far away from the absorbing inhomogeneous Pr ion transition (non-dispersive cavity). The crystal cavity is $\approx6\;$mm long, the reflectivity was specified to $R_1\approx R_2\approx 95\%$ and $Y_2SiO_5$ has a real index of refraction of $n\approx1.8$. The cavity mode spacing for this crystal is $\Delta \nu_{non-dispersive}\approx 13\;$GHz and the transmission peak linewidth is $\delta \nu_{non-dispersive}\approx 1\;$GHz.
A frequency scan across two cavity modes is shown in Fig. \ref{Readout}a. The reason for the comparatively large cavity linewidth (low finesse) could be due to improper matching to the spatial cavity mode \cite{Sabooni2013,Sabooni2012c}. We expect that the high frequency wing of the cavity modes is due to higher order transverse modes in the cavity. The small extra peak to the left of the modes is the cavity mode for the orthogonal polarization. The $Y_2SiO_5$ crystal is birefringent and cavity modes for the two polarizations along the two principal axes will be displaced relative to each other \cite{Sabooni2013a}. Both modes are seen because the input polarization has a small angle relative to the $D_2$ (principal) axes (see in Fig.\ref{setup}b).

Second, persistent spectral hole burning is employed to manipulate the ion absorption distribution in the inhomogeneously broadened ($\sim$ 9 GHz) $^3H_4-\!^1D_2$ transition of the $0.05\%$ doped $Pr^{3+}\,:Y_2SiO_5$ crystal. A frequency stabilized dye laser \cite{Drever1983,Leibrandt2011} at $\lambda_{vac}=605.976\;$nm is used to remove the absorbing ions within an $18\;$MHz wide spectral region (black trace, Fig. \ref{Readout}b) by a series of laser pulses. The laser pulses optically pump the ions to the excited state, from where they decay back to one of the hyperfine ground levels (see Fig. \ref{setup}a). A typical value for the ground state hyperfine population decay is $\sim$40 seconds, while it could be 40 minutes in a weak magnetic field ($\approx 0.01\;$T) \cite{Ohlsson2003}, and several weeks for other rare-earth-ion-doped crystals \cite{Konz2003}. A detailed description of the procedure for creating a spectral transmission window can be found in Ref. \cite{Amari2010}. An arbitrary-waveform-generator controlled double-pass acousto-optic modulator (AOM), which can tailor the light amplitude, phase and frequency, see Fig. \ref{setup}b, is used to form the necessary laser pulses. As is shown in Fig. \ref{Readout}b the modified ion absorption profile (black trace) has close to zero absorption within about 18 MHz. The absorption profile is taken in a part of the 6 mm crystal which is not reflection coated, i.e. outside the cavity. In this way the frequency resolved absorption measurement can be recorded without being affected by the cavity mode structure. Admittedly, this can only be seen as an estimate of the absorption structure inside the cavity. Previously we have measured the absorption, $\alpha L$, in a 1 mm crystal of the same dopant concentration to be approximately equal to 2. The left axis in Fig. \ref{Readout}b has been scaled to be consistent with this value. In order to verify that absorption and scattering losses are indeed small in the 18 MHz transmission window, we compared the transmission through the crystal (without the cavity) in the black-trace spectral transmission region in Fig. \ref{Readout}b, with the transmission through the crystal when the laser is detuned by about 4 nm from the 9 GHz inhomogeneous line center. It was not possible to detect a difference between the transmission in the 18 MHz transmission region and when the laser was detuned about 4 nm ($\sim$4 THz) away from the absorption line. From this we estimate that any absorption in the slow light transmission window is in any case below $<$ 0.1 dB/cm. The bulk loss in $Y_2SiO_5$ has previously been estimated to $<$ 0.003 dB/cm \cite{Goto2010}.
\par
Experimental line narrowing data is shown in Figs \ref{Readout}b and \ref{Readout}c. In Fig. \ref{Readout}b there are, five cavity transmission peaks within the $\approx18\;$MHz transparent spectral region. The spectral width of the cavity transmission is reduced from $\delta \nu_{non-dispersive}\approx 1\;$GHz in the non-dispersive cavity case (Fig. \ref{Readout}a) to $\delta \nu_{dispersive}\approx 600\;$kHz in the dispersive cavity case. In Fig. \ref{Readout}c the transmission window is about a MHz, yielding a steeper dispersion curve and the cavity line width is about 30 kHz, i.e. a reduction of about 30000 relative to the non-dispersive cavity line width. At the same time the cavity mode spacing has decreased from 13 GHz to about 220 kHz (factor of 60000).

\par
Eq. \ref{eq:Mode_Spacing} shows that it is possible to control the cavity mode spacing by controlling the group refractive index, $n_g(\nu)$.
The (real part of the) refractive index can be calculated from the ion absorption frequency distribution $\alpha(\nu)$ via the Kramers-Kronig relations. This will give the dispersion and group refractive index. The refractive index as calculated from the absorption (black line) is shown by the blue line in Fig. \ref{Readout}b. As shown in Fig. \ref{Readout} the cavity mode spacing and cavity line width can be controllably varied over several orders of magnitude by engineering the ion absorption frequency distribution.

\par
Although spectral hole-burning based slow light structures have been discussed previously \cite{Shakhmuratov2005,Lauro2009} and slow light structures in general have been analysed extensively, e.g. Ref. \cite{Boyd2011} and references therein, we think it could be relevant to discuss the properties and potential of the absorption structuring techniques shown here. Due to the cavity-linewidth narrowing, the $6\;$mm long cavity can have a longitudinal mode spacing of $\approx 220\;$kHz, which in vacuum would correspond to that of a $\approx 700\;$m long cavity! The transversal part of the mode however still behaves as that of a $6\;$mm long cavity, which means that it is possible to combine a narrow beam with fairly even beam size with a small cavity mode spacing. The longitudinal and transversal cavity mode behaviours are thus decoupled by more than four orders of magnitude.
\begin{figure}[ht]
    \includegraphics[width=8cm]{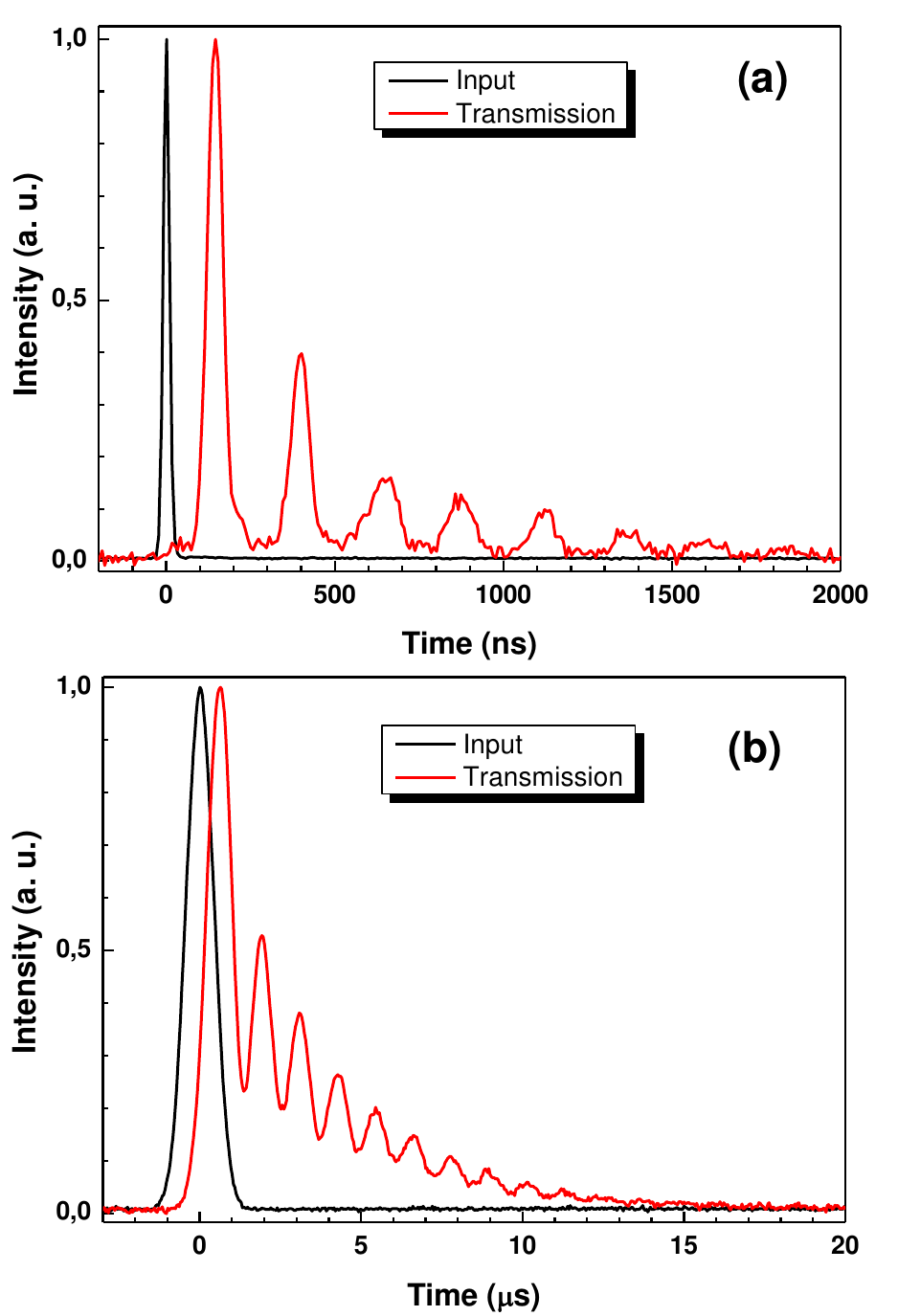}
    \caption{(Color online) Transmitted light to photo diode PD (a) When a 20 ns long pulse (black trace, centred at time t=0) is sent into the cavity in presence of the $\approx$18 MHz spectral hole. (b) When a 1 $\mu$s long pulse (black trace, centred at time t=0) is sent into the cavity in presence of the $\approx$3 MHz spectral hole. Both signals are normalized.}
    \label{Gaussian}
\end{figure}
\par
In order to further examine the cavity properties, a short pulse was sent into the cavity. The red solid trace in Fig. \ref{Gaussian}a shows the light transmitted to the photo diode, PD, for a 20 ns long input pulse (black dashed trace, centred at time t=0). With a 18 MHz slow light transmission window, the 6 m long pulse is compressed to about 2.5 mm and bounces back and forth in the cavity with some light leaking out after each round trip. Fig. \ref{Gaussian}b shows the corresponding data with a 3 MHz transmission window where the cavity round trip time is increased to well over a microsecond.  The group velocity, $v_g$, is approximately given by \cite{Walther2009a,Shakhmuratov2005}
\begin{equation} {\label{eq:group_velocity}}
v_g=\frac{2\pi\Gamma}{\alpha}
\end{equation}
where $\Gamma$ is the spectral width of the transmission window and $\alpha$ is the absorption coefficient immediately outside the spectral transmission window (i.e. the effective absorption depth of the transmission window). In general, \textit{e.g.} Ref. \cite{Lauro2009}, the time-bandwidth product, $TB$, for any slow light structure of this type of length $l$  approximately is
\begin{equation} {\label{eq:TB}}
TB=\frac{l}{v_g}\Gamma=\frac{\alpha l}{2\pi}
\end{equation}
Thus, the bandwidth is determined by the transmission window $\Gamma$, which readily can be changed by a new optical pumping sequence. The group velocity is then determined by the absorption coefficient of the surrounding structure which, for a given crystal, can most easily be changed by moving the transmission window to different positions within the inhomogeneous line. The further out from line center, higher the group velocity. Finally, the time delay is then set by the crystal length. It is noteworthy that all these parameters can be varied independently, which offers a good opportunity to test, for example, non-linear enhancement effects due to slow light in a well controlled environment. As an example a $TB\,\approx\,10$ for $\Gamma=1$ MHz, $\alpha\approx50$ cm$^{-1}$, and $l=$12 mm is reported in Ref. \cite{Zhang2012}.

\par
Potential applications of controlled spectral engineering of slow light structures in cavities will now be discussed. The fact that the lifetime of a light pulse in the cavity and the cavity Q-value increases by several orders of magnitude is interesting because whispering gallery mode rare earth crystal resonators with Q-values in the $10^6$ range have been demonstrated \cite{McAuslan2011b} and Q-values up to $10^{10}$ are predicted \cite{McAuslan2011b}. These numbers could potentially be enhanced by four orders of magnitude by modifying the absorption profiles by optical pumping as is done here. Also the loss rate due to scattering in the cavity, could be strongly reduced. For example, if the fractional scattering loss, $I_f$, for a pulse spending time, $t$, in the cavity is proportional to $1-e^{-\alpha_0l_0}$, where $\alpha_0$ is the scattering coefficient and $l_0$ is the length traversed by the pulse, then we obtain $I_f =1-e^{-\alpha_0l_0}= 1-e^{-\alpha_0v_gt}$, where $v_g$ is the group velocity. Thus if we would like to delay light by a given time, \textit{t}, a material with a low group velocity can strongly reduce scattering losses. In general these type of slow light effects can be interesting for solid state materials which as, compared to vacuum or gas, normally will experience larger scattering. %Through the slow light effect large delay and narrow linewidth is obtained even with lower reflectivity mirrors.

\par
It is also possible to dynamically tune the speed of light propagating through the material using an electric field. In non-centrosymmetric sites the rare earth ions have a permanent electric dipole moment which is different in the ground and excited state. This means that the ion transition frequency is shifted by an external electric field \cite{Macfarlane1994}, an effect which \textit{e.g.} has been used when creating high efficiency quantum memories in these materials \cite{Hedges2010} or to move the transmission window in frequency \cite{Beavan2013}. Electric fields can equally well be used to change the width of the spectral transmission window \cite{Persson2001} which changes the index of refraction. The group velocity can in this way readily be modulated at tens of MHz rates. As can be seen from Fig. \ref{Gaussian} this rate can be much faster than the time it takes for the pulse to propagate through the cavity. Thus it may be interesting to investigate the possibility to adiabatically tune the wavelength inside the cavity by modifying the refractive index as is done in photonic crystal cavities \textit{e.g.} Ref. \cite{Preble2007}, but now on a time scale which differs by six orders of magnitude.

\par
Low phase noise and very narrow line width laser systems have wide ranging applications as stable frequency references and sources. A passive laser line narrowing technique based on spectral hole burning (SHB) was demonstrated recently. Selective filtering of weak spectral components outside the laser centre frequency by transmitting the laser beam through a highly absorbing SHB material resulted in phase noise suppression by several tens of dB outside the central laser frequency \cite{Thiel2011}. Using the cavity line narrowing techniques presented in this paper, transmission windows narrower than the sub kHz homogeneous lines of the material can be obtained. Further, the region outside the transmission window would be non-transparent but light outside the center frequency is now reflected instead of being absorbed, which has several advantages. %By applying electric fields it should also be possible to dynamically change the properties of these filters.

\par
Spontaneous parametric down-conversion is widely used in quantum optics and quantum information science as a source for entangled photon pairs and as single photon sources. It is often desirable to generate entangled photons with a very narrow bandwidth, for example to match the bandwidth of an optical quantum memory or the bandwidth of an optical transition \cite{Clausen2011,Saglamyurek2011}. Photons with narrow bandwidth can be generated by cavity-enhanced spontaneous down conversion, where a periodically-poled non-linear crystal is enclosed inside a cavity \cite{Bao2008}. A rare-earth-ion-doped slow light cavity could be combined with a periodically-poled non-linear crystal. A single narrow cavity transmission peak could be created within the spectral transmission window of the rare-earth-ion-doped cavity crystal. Since the rare-earth-ion inhomogeneous absorption peak can be hundreds of gigahertz wide, this constitutes the only transparent region. %The exact dispersion could potentially be fine-tuned by applying an electric field, which will Stark shift the resonance frequency of the dopant ions.

\par
The present cavity has the unusual property of supporting several modes with equal wavelength but with different frequencies. From Eq. \ref{eq:cmodes} it can be seen that for any mode number, \textit{m}, mathematically there is an infinite number of combinations of  $\nu$ and $n$ which will fulfil the equation. The present cavity indeed has more than one mode supported for certain mode numbers. For instance, in Fig. \ref{Readout}b, the two first(last) modes, which are about 4 MHz apart, have the same mode number as the two cavity modes just below(above) the rare earth absorption line, which are separated by about 13 GHz.

%\section{Conclusion}
\par
In conclusion we have shown more than 4 orders of magnitude cavity linewidth narrowing caused by off-resonant interaction with praseodymium ions doped in an inorganic crystal. Several cavity modes are shown within the 18 MHz slow light transmission window. The crystal can readily be reprogrammed by optical pumping or dynamically tuned by external electric fields to yield other cavity line widths or light group velocities. It is suggested that the combination of slow light structures and whispering galley modes in rare earth crystal resonators might give exceptionally high cavity Q values.

%The inhomogeneous broadened praseodymium absorption is removed from a $18\;$MHz wide spectral region. The $6\;$mm crystal cavity doped with Praseodymium thus acts like a $40\;$m long cavity.

%\begin{acknowledgments}
\par
This work was supported by the Swedish Research Council, the Knut \& Alice Wallenberg Foundation, the Crafoord Foundation, the EC FP7 Contract No. 247743 (QuRep). The research leading to these results also received funding from the People Programme (Marie Curie Actions) of the European Union's Seventh Framework Programme FP7 (2007-2013) under REA grant agreement no. 287252(Marie Curie Action). Finally, we are grateful to Dr. Mikael Afzelius for several valuable discussions.

\bibliography{C:/MAHMOOD/PAPERS/Main_Ref}
\end{document}